\documentclass[usegraphicx,usenatbib]{mn2e}

\title[Seyfert galaxies that are undergoing merging]
{Seyfert galaxies that are undergoing merging but appear non-interacting.}

\author[A. A. Smirnova et al.]{A. A. Smirnova\thanks{E-mail:
ssmirnova@gmail.com}, A. V. Moiseev, V. L. Afanasiev\\ Special Astrophysical Observatory, Nizhnij Arkhyz 369169, Russia}

\begin{document}

\date{Accepted 2010 May 31 . Received 2010 May 24 ; in original form 2010 April 21}

\pagerange{} \pubyear{2010}
\maketitle

\label{firstpage}

\begin{abstract}
We present new broad-band optical images of some merging Seyfert galaxies that were earlier considered to be non-interacting objects. On our deep images obtained at the Russian 6-m telescope we have detected elongated tidal envelopes belonging to satellites debris with a surface $R$-band brightness about $25-26.5\,\mbox{mag}\,\mbox{arcsec}^{-2}$. These structures are invisible in Sloan Digital Sky Survey (SDSS) pictures because of their photometric limit. We  found that 35 per cent of the sample of isolated galaxies has undergone merging during the last $0.5-1$ Gyr. Our results suggest that statistic studies based on popular imaging surveys (SDSS or Second Palomar Observatory Sky Survey (POSS-II)) can lead to underestimation of the fraction of minor mergers among  galaxies with active nuclei (AGN). This fact impacts on statistics and must be taken into consideration when finding connection between minor/major merging or interactions and nucleus activity.
\end{abstract}


\section{Introduction}

Theoretical studies show that galaxy interactions can bring gas from
the outer region of the disc toward the nucleus, and also produce
a burst of star formation and/or trigger nucleus activity \citep{mih96,cat05,spr05}. Therefore, statistical studies of interacting and non-interacting AGN galaxies are very important.

Many authors have tried to find a correlation between the presence
of an AGN in a galaxy and its environment: the existence
of companions or traces of interaction \citep{Dahari1985,DeRobertis1998,Schmitt2001,Knapen2005}. However, statistically
significant correlation of the properties of the host galaxies with
the activity wasn't found in any of the listed papers. \citet{Kauf04} and \citet{HR04} compiled various samples
of galaxies with active and normal (quiescent) nuclei from the
Sloan Digital Sky Survey (SDSS) and compared their morphological
properties. Again, no statistically significant differences
between galaxies with active and normal nuclei were found.

 In the recent papers \citet{Kuo08} and \citet{Tang08}, using HI observations of normal galaxies
and galaxies with AGN, demonstrated the absolute prevalence of tidal interactions among local Seyfert galaxies with relatively high luminosity. They conclude that \textit{"the dramatic contrast in the incidence of HI disturbances between active and inactive galaxy samples strongly implicates tidal interactions in initiating events that lead to luminous Seyfert activity in a large fraction of local disc galaxies"}. Interestingly, only small fraction of their sample galaxies are visibly disturbed in optical starlight in the SDSS images. \citet{Kuo08} explain this dramatic difference between HI gas and optics by a longer time of dynamical relaxation for the outer regions of HI disc.

Some authors (see, for example, short review of \citet{SB08} and references therein) propose that the interaction is only a first step in the process of gas inward redistribution. There is a significant time  delay between the interaction and the phase at which the black hole is able to accrete (Li et al. 2008). In this case, it might be more important to look for the signs of past interaction or merging (tidal envelopes, tails etc.), not only for galaxy-satellite pair.

Data from the SDSS, widely used for statistical studies  can't ensure a sufficient photometric limit for the detection of faint tidal structures around galaxies.  Only deep imaging can solve this problem. A good illustration is the project undertaking systematic deep images of stellar tidal streams from a sample of $\sim$ 50 nearby Milky-Way-like spiral galaxies within 5 Mpc \citep{md08,md10}. They have discovered faint loop-like features in the outer regions of several spiral galaxies that appear to be undisturbed in high surface brightness optical images but are warped in HI maps.

 In this paper, we present deep images of some Seyfert galaxies observed with the 6-m Big Azimuthal Telescope (BTA) of the Special Astrophysical Observatory of the Russian Academy of Sciences (SAO RAS).

 The paper is organized as follow: in Section~\ref{obs} we present details  of the observations, in Section~\ref{ima} we compare our data with the POSS-II and SDSS images and in Section~\ref{concl} we give a short conclusion.

\begin{center}
\begin{table}
\caption{Sample of isolated galaxies.}
\label{sample}
\begin{tabular}{lllll}
\hline
Object Name & Sy Type & Isolation & New Shells & \\
\hline
  Mrk  78   &    2    &   abs.    &     --    &   \\
  Mrk 198   &    2    &   abs.    &     +     &   \\
  Mrk 291   &    1    &   abs.    &     --    &   \\
  Mrk 315   &   1.5   &  moder.   &     +     &   \\
  Mrk 334   &   1.8   &   abs.    &     +     &   \\
  Mrk 335   &   1.2   &   abs.    &     --    &   \\
  Mrk 359   &   1.5   &   abs.    &     --    &   \\
  Mrk 493   &   NLS1  &   abs.    &     --    &   \\
  Mrk 516   &   1.8   &   abs.    &     +     &   \\
  Mrk 543   &   1.5   &   abs.    &     +     &   \\
  Mrk 766   &   NLS1  &   abs.    &     --    &   \\
  Mrk 885   &   1.5   &   abs.    &     --    &   \\
  Mrk 896   &   NLS1  &   abs.    &     --    &   \\
  Mrk 917   &    2    &   abs.    &     +     &   \\
  Mrk 1058  &    2    &   abs.    &     --    &   \\
  Mrk 1066  &    2    &   abs.    &     +     &   \\
  Mrk 1073  &    2    &  moder.   &     --    &   \\
  Mrk 1179  &   1.9   &   abs.    &     --    &   \\
  NGC 2110  &    2    &  moder.   &     --    &   \\
  UGC 3478  &   1.2   &   abs.    &     --    &   \\
\hline
\end{tabular}
\end{table}
\end{center}

\begin{center}
\begin{table}
\caption{Log of observations.}
\label{speclog}
\begin{tabular}{lllll}
\hline
Object Name &   Date            & T$_{exp}$ & Filter &  Seeing   \\
            &                   &   (sec)   &        &  (arcsec)  \\
\hline
\multicolumn{5}{c}{\it Isolated Galaxies}\\
  Mrk 198   & 2005 December 26  &  720 &  $V$  &  2.0  \\
            &                   &  480 &  $R$  &  2.0  \\
  Mrk 315   & 2004 September 07 &  1320&  $V$  &  1.5  \\
            &                   &  660 &  $R$  &  1.5  \\
  Mrk 334   & 2006 October 23   &  600 &  $R$  &  1.4  \\
  Mrk 516   & 2008 October 28   &  600 &  $V$  &  1.4  \\
  Mrk 543   & 2008 October 28   &  720 &  $R$  &  1.5  \\
  Mrk 917   & 2008 September 03 &  450 &  $R$  &  1.4  \\
  Mrk 1066  & 2008 October 22   &  900 &  $R$  &  1.5  \\
\multicolumn{5}{c}{\it Non-Isolated Galaxies}\\
  Mrk  993  & 2008 October 27   &  600 &  $R$  &  1.2  \\
  Mrk 1146  & 2008 October 27   &  600 &  $R$  &  1.3  \\
  Mrk 1469  & 2005 May     19   & 1000 &  $R$  &  1.2  \\
\hline
\end{tabular}
\end{table}
\end{center}

\section{Sample selection and observations}\label{obs}

\subsection{The sample of isolated galaxies.}

During 1998-2009, several small samples of nearby $(z<0.04)$ Seyfert galaxies were studied using a 3D spectroscopy technique used by the 6-m (BTA) telescope \footnote{In different years the prime investigators of the projects were V. Afanasiev, G. Richter, P. Rafannelli, S. Ciroi, A. Moiseev and  A. Smirnova.}. These observations were aimed at a detailed investigation of the gas ionization properties as well as stellar and gas  kinematics in the target galaxies, in order to find a possible mechanism of AGN fueling related to jet/clouds interaction etc. The  results for several individual galaxies were published \citep[see, for instance,][]{Ciroi01,Ciroi05,Radovich05,Smir06,Smir10}.
In order to improve our knowledge about targets for which  3D spectroscopic data were available, a deep imaging observation programme was started. Examination of the external  parts of galaxies  can help us to understand not only galaxy morphology but also the merging history. Past interactions can play an important role in supplying AGN with gas, but deep  images of selected objects are not available in open data bases and literature. We now collect deep images of more than 30 galaxies observed  with BTA. Of special interest are  galaxies described earlier as `isolated' and  `non-interacting'. Using NED database\footnote{http://nedwww.ipac.caltech.edu/}, we check the isolation of galaxies by applying the criteria proposed by \citet{Schmitt2001}: (1) the distance from the main galaxy to the possible companion must be not smaller than $5\times D_{25}$; (2) the difference in brightness between them must be no smaller than 3 mag; (3) the difference in radial velocities must be no smaller than 1000 km s$^{-1}$. The final sub-sample of isolated galaxies contains 20 objects listed in Table~\ref{sample}.  This sub-sample of isolated galaxies, selected  from the list of objects  observed with 3D spectroscopy, will hereinafter be referred to as `the sample'. In Table~1 we divided  the sample into two parts: an absolutely isolated one and a moderately isolated one. Absolutely isolated objects satisfy all the three Schmitt criteria, while moderately isolated galaxies satisfy the brightness and distance criteria without any information about velocity. We also include in our paper three galaxies (Mrk~993, Mrk~1146, Mrk~1469) with distant satellites: new outer structures were found in their discs that had been unknown before.

\subsection{BTA deep imaging}
The observations were made with the Spectral Camera with Optical Reducer for Photometrical and Interferometrical Observations (SCORPIO, \citet{Afan05}) mounted at the BTA prime focus. We obtained images  in the Johnson-Causins $V$, $R$ bands with a scale of 0.35 arcsec per pixel in a field $6.1\times6.1$ arcmin$^2$; the detector was a CCD EEV42-40 ($2048 \times 2048$ pixels). All the targets listed in Table~\ref{sample} were observed, but below we present only those images of galaxies where low brightness extended features were found (see Table~\ref{speclog} for the observational log). Data reduction was performed in IDL and included a number of standard procedures: bias subtraction, flat fielding, and cosmic ray particle hits removal.  The photometric calibration was  based on  the aperture photometry data listed in the HyperLeda data base\footnote{http://leda.univ-lyon1.fr/}. The surface brightness magnitude limit in the $R$-band for the galaxies listed in  Table~\ref{speclog} is $26.0\pm0.3\,\mbox{mag}\,\mbox{arcsec}^{-2}$ for the signal-to-noise ratio $S/N=3$ in an individual pixel.

\subsection{SDSS and POSS-II data}
All the observed galaxies have been cross-matched with SDSS Data Release 7  \citep[][ SDSS-DR7]{Aba09}.
We used calibrated  FITS-files from SDSS DR7 in all bands ($u$, $g$, $r$, $i$, $z$) and in order to increase signal-to-noise ratio we co-added images in five photometric bands for each galaxy. SDSS DR7 has no data for three galaxies: Mrk~334, Mrk~516 and Mrk~543. For these objects we used digitized red plates of the Second Palomar Observatory Sky Survey (POSS-II).

\begin{figure*}
\begin{center}
\centerline{
\includegraphics[height=15 cm]{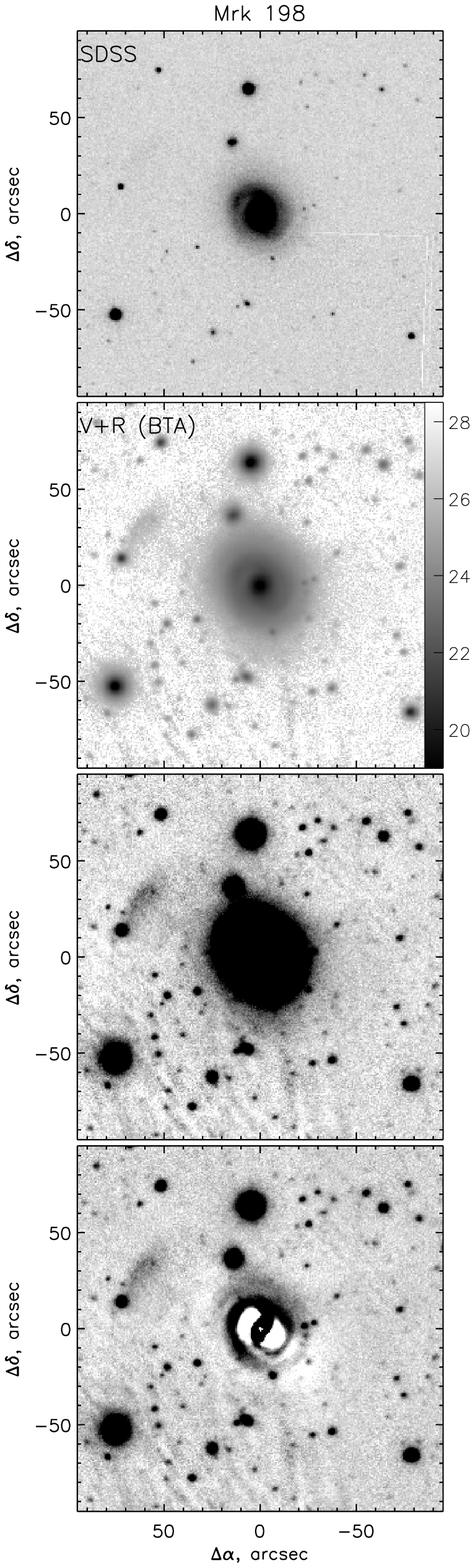}
\includegraphics[height=15 cm ]{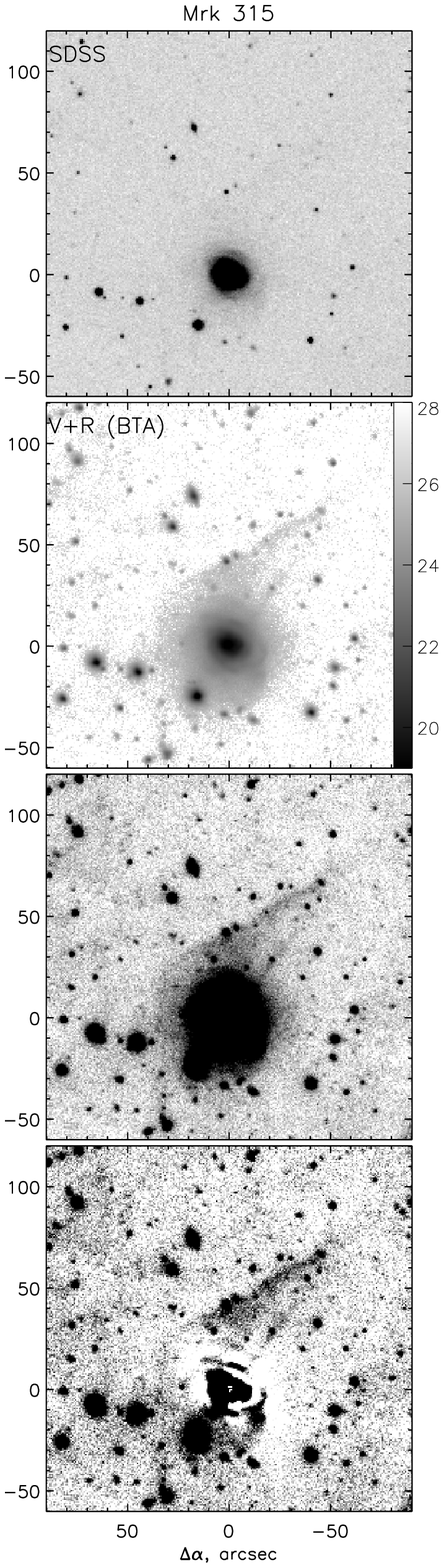}
\includegraphics[height=15 cm]{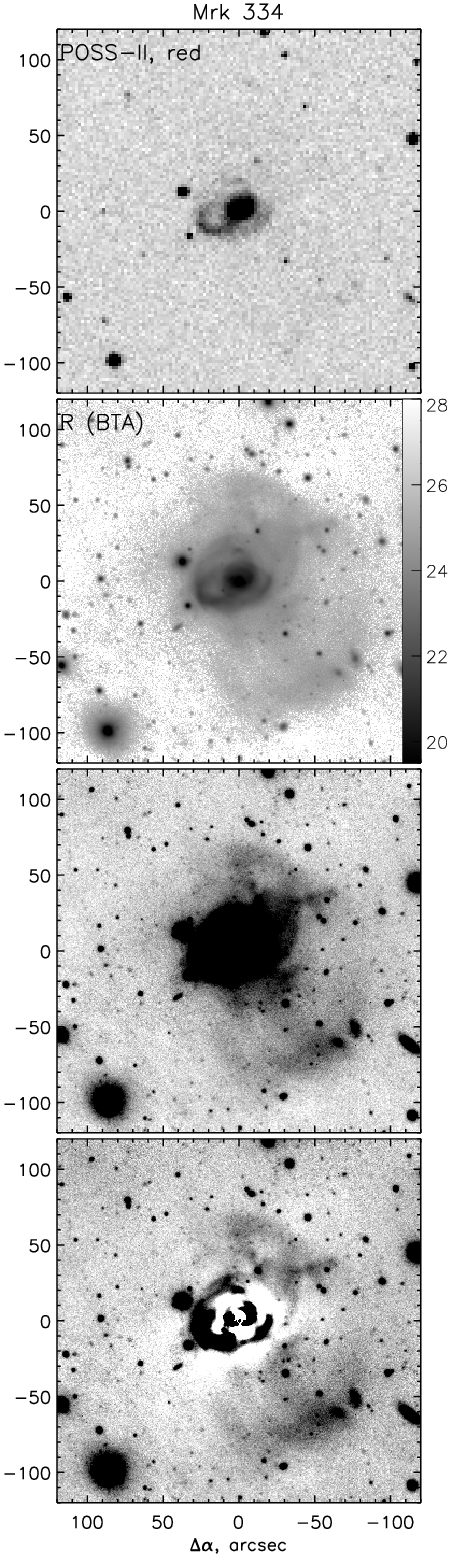}
\includegraphics[height=15 cm ]{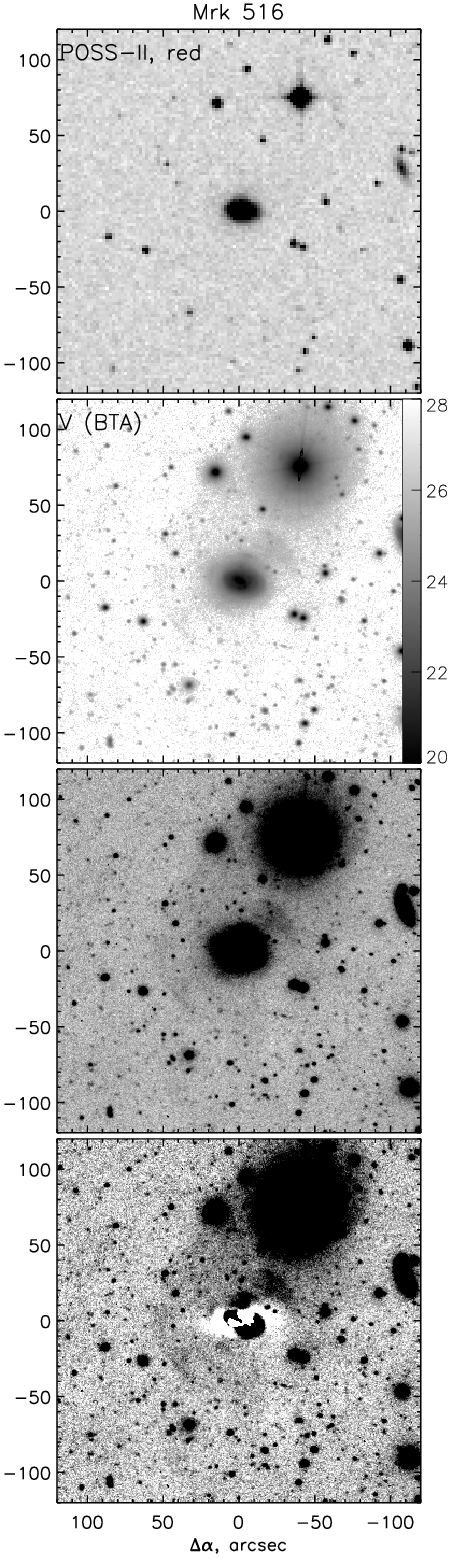}
}
\caption{Images of galaxies Mrk~198, Mrk~315, Mrk~334, and Mrk 516. From top to bottom: an archival (SDSS-total or POSS-2 red) image; BTA deep image in magnitudes, where the band ($V$,$R$ or $V+R$) is labelled and the scale box shows the surface brightness in $\mbox{mag}\,\mbox{arcsec}^{-2}$; BTA deep image on linear scale ; a residual image after 2D model subtraction.}\label{fig_01}
\end{center}
\end{figure*}

\begin{figure*}
\begin{center}
\centerline{
\includegraphics[height=15 cm]{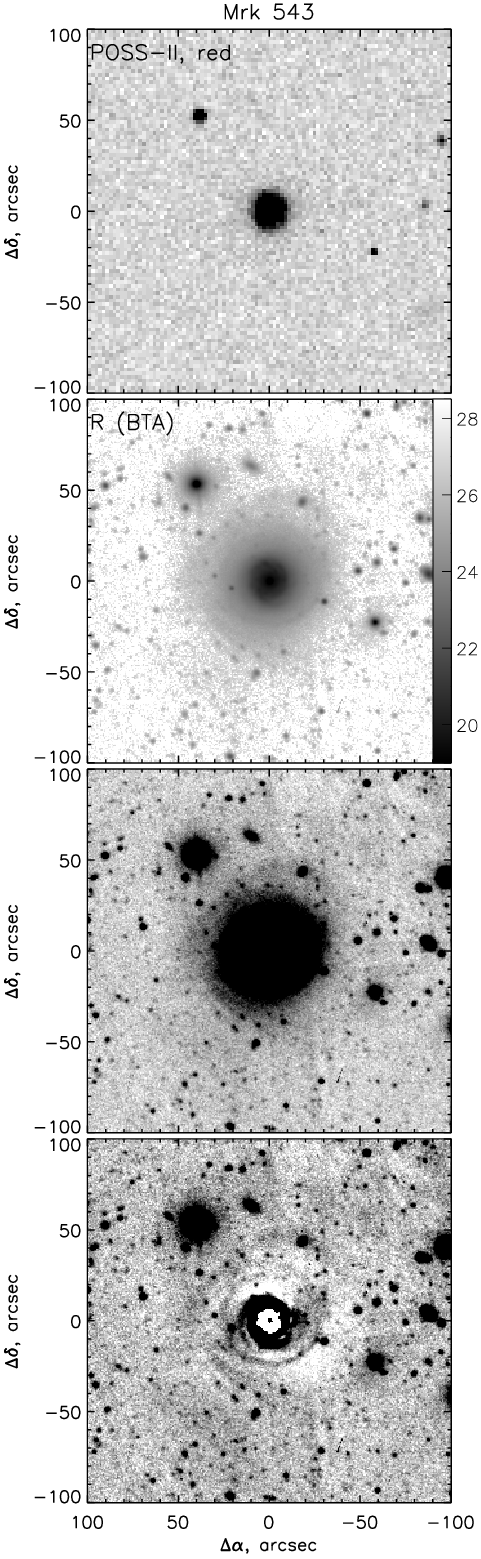}
\includegraphics[height=15 cm ]{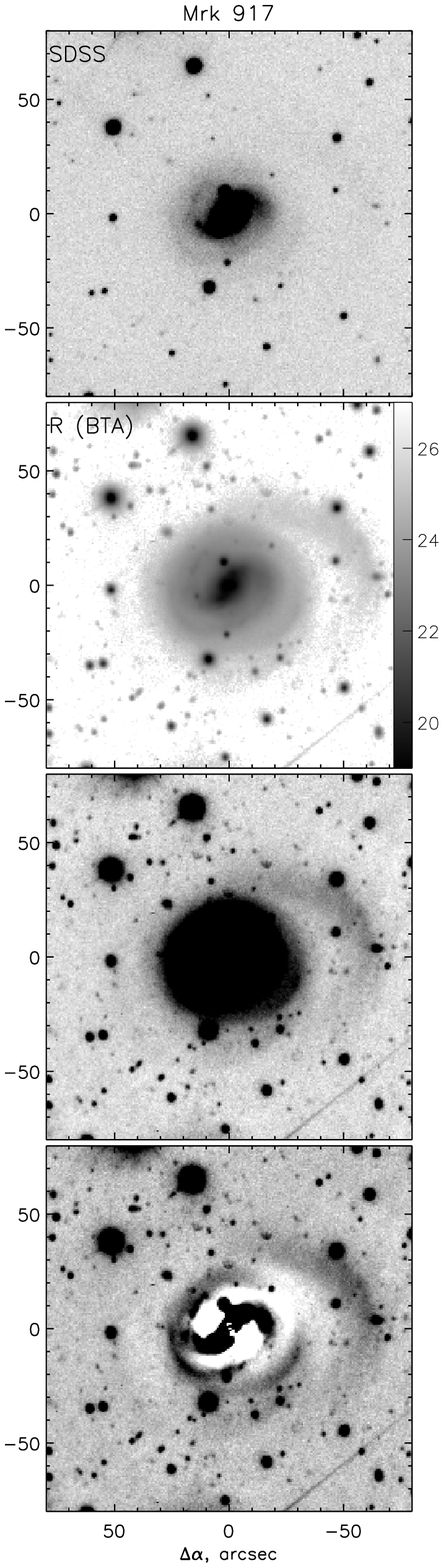}
\includegraphics[height=15 cm]{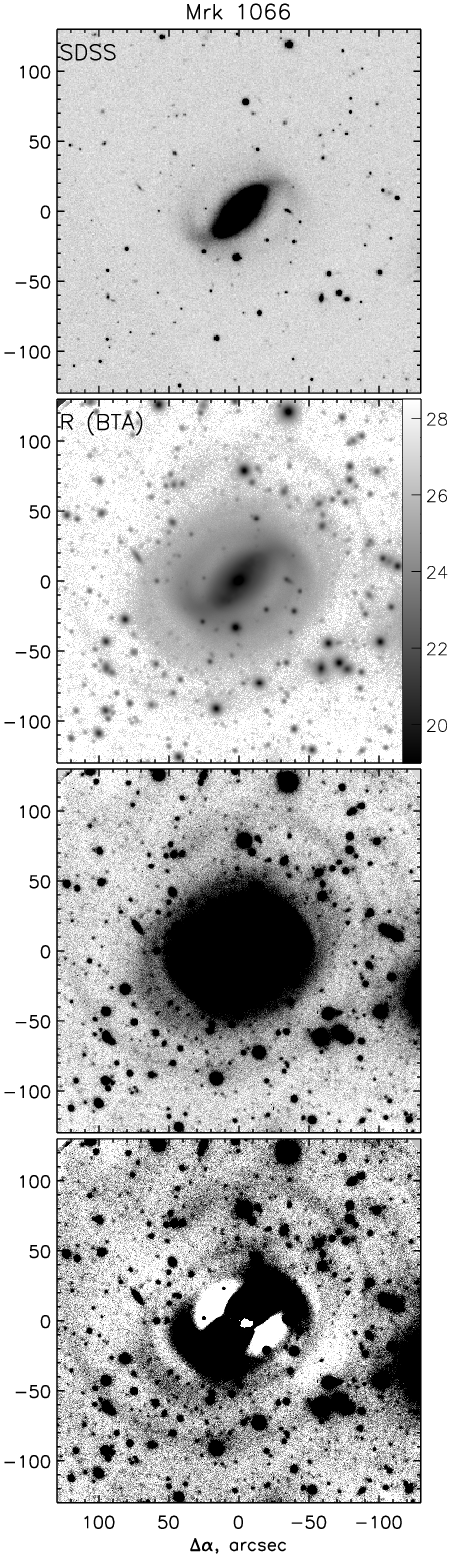}
}
\caption{Same as in Fig.~\ref{fig_01} but for galaxies Mrk~543, Mrk~917 and Mrk~1066.}\label{fig_02}
\end{center}
\end{figure*}

\begin{figure*}
\begin{center}
\centerline{
\includegraphics[height=15 cm ]{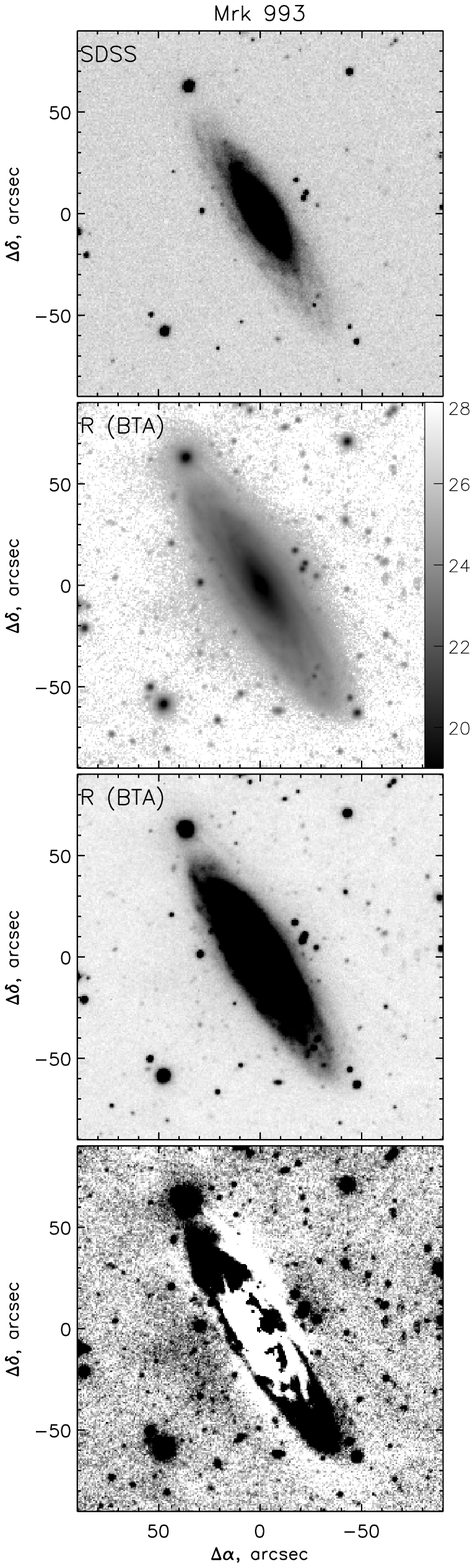}
\includegraphics[height=15 cm]{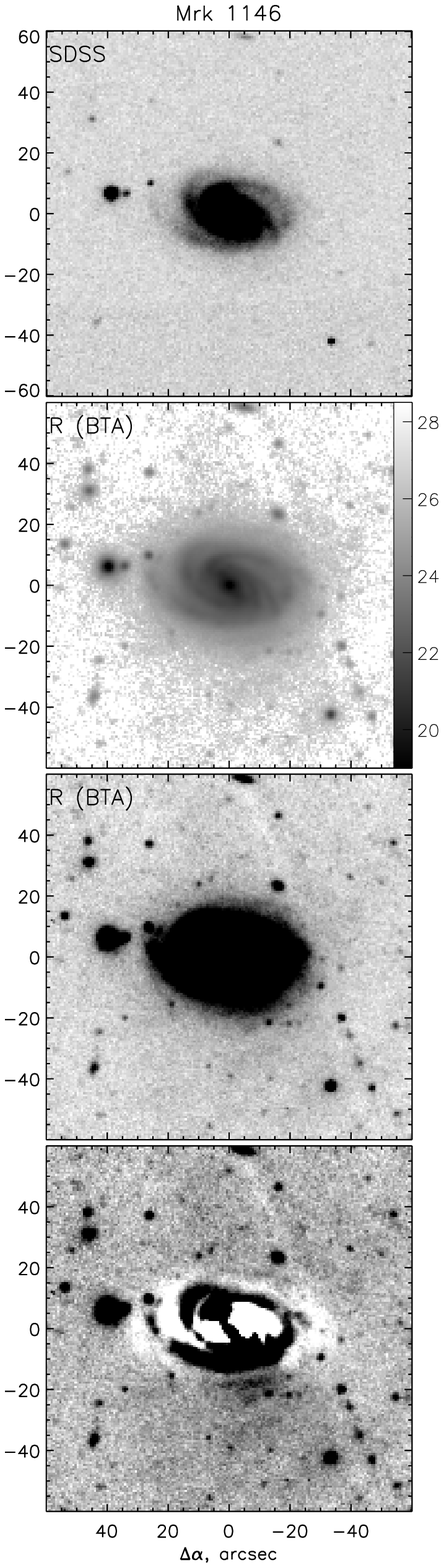}
\includegraphics[height=15 cm ]{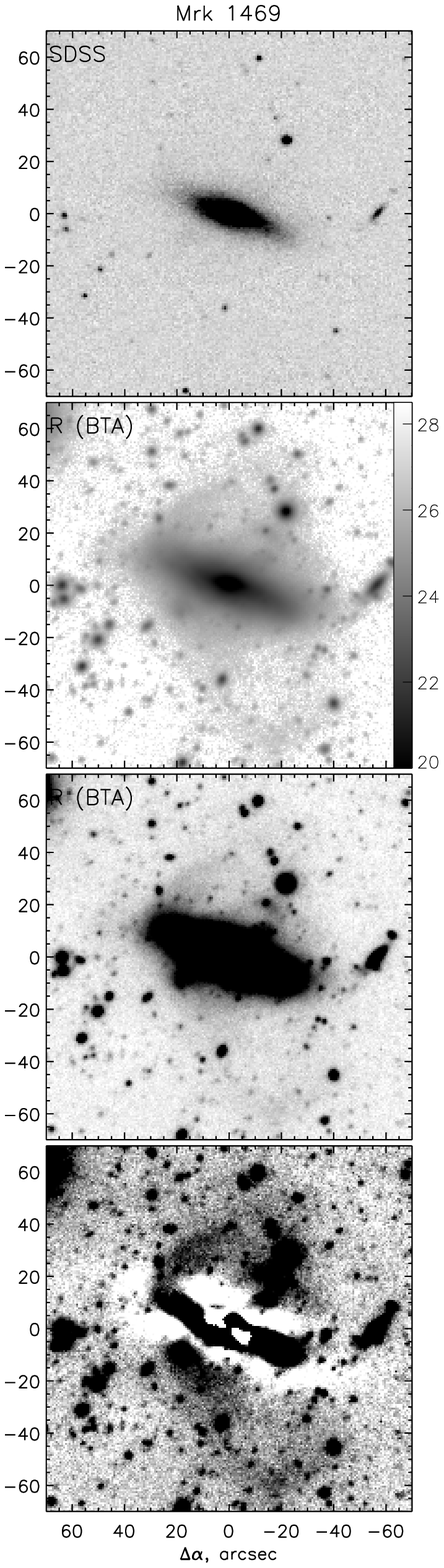}
}
\caption{Same as in Fig.~\ref{fig_01} but for non-isolated galaxies Mrk~993, Mrk~1146 and Mrk~1469.}\label{fig_03}
\end{center}
\end{figure*}

\section{BTA data vs SDSS images: individual objects}\label{ima}

In this section we compare BTA direct images with the SDSS ones for individual galaxies. In this paper we only consider objects from the sample where tidal structures were detected on the BTA deep images. In seven out of twenty isolated Seyfert galaxies from Table~\ref{sample},  faint envelopes were detected.  To study the brightness distribution in the filaments, we removed the axisymmetric components of galaxy bulge and disc. To decompose the image into components, we used an iterative method of constructing 2D models discussed by \citet{Mois04,Ciroi05}. As a result, for every object Figures~\ref{fig_01}-\ref{fig_03} show the deep BTA image, co-add SDSS images in five photometric bands and brightness distribution after subtracting a model consisting of a disc and a bulge. This procedure increases a contrast of faint extended non-symmetric morphological structures.

The low-brightness envelopes in the two systems have previously been described in our papers(see below references to  Mrk~315 and Mrk~334). However we also present  their images here for  comparison with other galaxies to make the results homogeneous.

\subsection{New structures in  isolated galaxies.}\label{main}

\textbf{\emph{Mrk~198}}: Faint arc-like condensation is seen at 60-70 arcsec north-east from the nucleus Mrk~198 on our deep image (Fig.~\ref{fig_01}). This feature is absent in SDSS data.

\textbf{\emph{Mrk~315}}: At a first glance the SDSS morphology of Mrk~315 does not show signs of past or ongoing interaction.
However, deep images reveal two filaments in the surroundings Mrk~315 (see Fig.~\ref{fig_01}). Detailed study of this galaxy \citep{Ciroi05} shows that these two filaments are debris of two dwarf companions. Specifically, one sank into the main galaxy and gave rise to a minor merger event also forming twister tidal filaments NE from the main galaxy, while the other one passed close to Mrk~315 in a kind of fly-by.

\textbf{\emph{Mrk~334}}: In the POSS-II image an asymmetric spiral arm resembling a tidal tail east of the nucleus is present. It is clear even at a first glance  at the deep images of the galaxy that the tidal arm is the brightest part of a vast system of shells and lower  surface brightness filaments (Fig.~\ref{fig_01}). These shells  have sharp outer edges located about  70 arcsec north-west and  100 arcsec south-west of the nucleus. The  analysis of this image and various spectral data \citep{Smir10} leads us to conclude that  Mrk 334 is in the process of a merger with a companion that has already been almost completely disrupted by tidal forces. The shells are the debris of this satellite and they mark the satellite's movement around Mrk~334. Brightness estimation for the tidal filaments yields a mass ratio of  $1/5$ to $1/3$ for the interacting galaxies, which is close to the adopted boundary between minor and major merging \citep{Smir10}.

\textbf{\emph{Mrk~516}}: According to HST observations \citep{Deo06}, the nuclear dust morphology is chaotic, extensive star formation is taking place, and the nuclear region shows two bright nuclei. However, outer parts of this galaxy look undisrupted in the POSS-II plate. The deep image shows faint diffuse cloud-like shells around the main body of Mrk~516 at distances of 10 to 40 arcsec (see Fig.~\ref{fig_01}).  All these peculiarities can be caused by merging with satellites that have been completely destroyed in Mrk 516$^,$s gravitational field. Although this object satisfies the isolation criterion adopted in this paper, POSS-II plates, as  well as our deep images, show a galaxy at a distance $\sim$100 arcsec east from Mrk~516. The difference in brightness between these two objects is $\Delta m   = 3$ mag by our calculations. Our coordinate estimation gives $\alpha_{2000} = 21^h56^m15.1^s$, $\delta_{2000} =  +07^{\circ}22^{'}27^{''}$. There is not any object near this position in the NED data base.  It is unclear whether it is a field galaxy or an Mrk~516 satellite until its redshift is measured.

\textbf{\emph{Mrk~543}}: This galaxy shows dust on a large scale. Multiple spiral arms are seen, littered with star-forming regions. However, the central 1 –- 2 kpc appear to be smooth and devoid of dust \citep{Deo06}. The outer regions of Mrk~543 demonstrate a diffuse envelope at a distance of up to 50 arcsec. On  subtraction of the 2D exponential  disc $+$ bulge model, this envelope transforms itself into a regular system of shells or ripples (see Fig.~\ref{fig_02}). These weak  structural features  appear similar to  low-contrast shells or spiral arm fragments found by \citet{Zasov08} in the NGC~6340 lenticular galaxy. Using  photometric and kinematic analysis, \citet{Chil2009} show that properties of NGC~6340 can be explained as the result of both major and minor events. Therefore, we believe that a similar structure observed in Mrk~543 also evidences a close interaction with another galaxy.

\textbf{\emph{Mrk~917}} is a nearby isolated Sy2 galaxy \citep{Ciroi99}.  This galaxy shows the distorted pattern of the velocity field of the gas and high star formation rate in a region close to the nucleus \citep{Ciroi01}.  Our deep images (Fig.~\ref{fig_02}) demonstrate an arm-like structure seen in the outer parts of the galaxy. The brightest of these filaments at  50 - 60 arcsec NW from the nucleus looks like a  tidal stream that arose from possible merging with an unknown companion. This interaction perturbs the ionized gas  velocity  field and can also trigger a star formation  burst in Mrk~917. The morphology of the outer parts of Mrk~917 looks similar to filaments in NGC6104, which are caused by merging with a satellite (for details see \citet{Smir06}).

\textbf{\emph{Mrk~1066}} is an isolated inclined spiral galaxy. Although it has no satellite or sign of interaction, very high color excesses on opposite sides of the nucleus are seen in Hubble Space Telescope Near Infrared Camera and Multi-Object Spectrometer (HST NICMOS) maps \citep{Regan99}. In spite of the regular spiral shape in the SDSS-image, our deep $R$-image reveals a vast system of shells or ripples in the outer parts of Mrk~1066 disc (Fig.~\ref{fig_02}). They are located at a distance of 50 to 100 arcsec. But it is not known if these ripples are associated with one or several different merger events. However, all peculiarities indicate a recent merging of Mrk~1066 with a satellite (or satellites): traces of this process can be observed as dust peculiarities in the body of galaxy and as tidal ripples in the outer parts of Mrk~1066.

\begin{center}
\begin{table}
\caption{List of non-isolated galaxies.}
\label{sample1}
\begin{tabular}{lllll}
\hline
Object Name & Sy Type & Isolation & New Structures & \\
\hline
  Mrk  993  &  2/1.5  &  moder.   &  galactic wind & \\
  Mrk 1146  &    1    &  dist sat &     shells    &   \\
  Mrk 1469  &   1.8   &  dist sat & polar ring + shells &   \\
\hline
\end{tabular}
\end{table}
\end{center}

\subsection{New structures in  non-isolated galaxies.}\label{add}
 In this paper, we present three additional objects that have distant satellites and previously  unknown faint structures in the outer parts of their discs (see Table~\ref{sample1}).

\textbf{\emph{Mrk~993}}: The residual deep image  reveals faint spacious structures spread along the galaxy's minor axis. These structures extend up to 30 arcsec East and 60 arcsec West of the Mrk~993 nucleus (Fig.~\ref{fig_03}). The shape of these features mostly resembles a galactic wind (superwind) caused by a powerful starburst \citep[see][for review]{Veilleux2005}. In this case,  the bright emission lines H$\alpha+\mbox{[N\,{\sc ii}]}$ in a superwind cone  produced a contamination to the  $R$-filter image. New narrow-band and deep spectral observations are needed for confirmation of our supposition.

\textbf{\emph{Mrk~1146}} has a satellite named SDSS J004728.47+144502.1 located at a distance of 3.6 arcmin (about 4.6xD$_{25}$). According to SDSS data, this is an undisturbed galaxy, but a deep $R$-image shows a faint plume extended 20 arcsec south from the nucleus (Fig.~\ref{fig_03}).

\textbf{\emph{Mrk~1469}}: this high-inclined  galaxy has a satellite named  SDSS J121617.69+505020.4  located at a distance of 1.9 arcmin (about 2xD$_{25}$). The $R$-band deep image of outer isophotes reveals a weak symmetric feature extended along a $\sim115$ deg position angle at a distances of about $20-30$ arcsec from the nucleus. It resembles structures observed in some polar ring galaxies, such as
for instance ESO 415-G26 \citep[see Fig7b in][]{Whitmore1987}. According to the current conception,  polar rings are formed  via galaxy mergers or accretion of matter of the companion \citep[see for review][]{bc03}. However, in contrast with a `true' polar ring, the  possible tidal stream in Mrk~1469 seems to be unstable because its plane is not orthogonal to the disc of the main galaxy but is low inclined  (about $45$ deg). Therefore, it is a short-living structure formed during the last (about $1-2$ Gyr) merging event. Moreover, on the residual image  (Fig.~\ref{fig_03}) we detected a vast system of low-brightness shells  at different spatial scales (up to 70 arcsec). These tidal shells similar to the inner inclined structure,  seem to be formed at the same event of interaction. At the same time the SDSS image dosen't show any features in the outer parts of the galaxy.

\section{Conclusion}\label{concl}

We analysed broad-band optical images of some merging Seyfert galaxies that had been previously considered to be non-interacting. In the deep images taken by the Russian 6-m telescope, we have found the following:

\begin{enumerate}
\item elongated tidal structures in 7 strong isolated Seyfert galaxies that had been previously considered to be non-interacting objects (Mrk~198, Mrk~315, Mrk~334, Mrk~516, Mrk~543, Mrk~917, Mrk~1066)

\item  previously unknown traces of past merging in 3 Seyfert galaxies, that have satellites but appear like undisturbed objects (Mrk~993, Mrk~1146, Mrk~1469)

\end{enumerate}

We compare our observation data with the SDSS and POSS-II images and can conclude that these faint structures are not seen in SDSS and POSS-II even when we add SDSS images in all photometric bands for each galaxy. We suggest that the Sloan Digital Sky Survey  data widely used for statistical studies are unable to ensure a sufficient photometric limit for detecting faint tidal structures around galaxies.

Our sample consists of 20 strongly isolated Seyfert galaxies, 7 of which (i.e. $35\pm11$ per cent at $1\sigma$ level) were found to have traces of past merging. The life time of the observed tidal features is approximately equal to the dynamical time of the corresponding radii (orbital time-scale). Our estimations provide $t_{dyn} \approx 0.4 - 1.2$ Gyr for these galaxies, if maximal rotation velocity is assumed to be 200 km s$^{-1}$. Therefore, the recent merging events could trigger nuclear activity in these galaxies. Since the interaction can be a first step in sending gas inwards, it is very important to find out the role of merging in initiating AGN activity.  In this sense, the new paper of \citet{md10} is of great interest: the authors obtained ultra deep images (up to 27 mag arcsec$^{-2}$) of several isolated spiral galaxies in the Local Volume. As a result, six previously undetected giant stellar structures around selected galaxies were discovered. We suggest that only a survey of deep images can ensure a comprehensive examination of the interecting/merging history in AGN galaxies. For a thorough study of the role of merging in initiating AGN-activity, a matched sample of non-Seyfert galaxies at a similar photometrical depth is needed before any conclusions can be drawn.

In most cases, when we analyse galactic morphology and kinematics in detail (see, for example, Mrk~315, Mrk~334), we can find remnants of the central part of a disrupted galaxy-satellite. In other cases, however, we can only assume that the merging involved a normal dwarf galaxy (possiby with a dark halo or `dark galaxy' \citep{ikar06}). A detailed morphological and kinematic study of disrupted isolated galaxies  without visible satellites  can improve our understanding of galaxies evolution and properties of dark matter.

It would be very interesting to carry out a deep optical imaging of large samples of Seyfert and non-Seyfert galaxies. Thereby we could investigate the merging histories of active and non-active galaxies, compare them and find out if there is any difference between these objects or not. In any case, it would be very useful to retrace the dependance of the stage of AGN activity on the stage of merging.

\section*{Acknowledgments}

This work based on observations carried out at the 6-m telescope
of the Special Astrophysical Observatory of the Russian Academy of
Sciences, operated under the financial support of the Science
Department of Russia (registration number 01-43). Funding for the SDSS and SDSS-II has been provided by the Alfred P. Sloan Foundation, the Participating Institutions, the National Science Foundation, the US Department of Energy, the National Aeronautics and Space Administration, the Japanese Monbukagakusho, the Max Planck Society and the Higher Education Funding Council for England. The SDSS web site is http://www.sdss.org/. This research
has made use of the NASA/IPAC Extragalactic Database (NED) which
is operated by the Jet Propulsion Laboratory, California Institute
of Technology, under contract with the National Aeronautics and
Space Administration.  This work was partly supported by the
Russian Foundation for Basic Research (project 09-02-00870). We are grateful to our referee Jeremy Lim for his constructive comments, which helped us to improve the paper. We also thank Olga Smirnova for grammar correction.

\end{document}